\documentclass[runningheads]{llncs}
\usepackage[T1]{fontenc}
\usepackage{graphicx}
\usepackage{hyperref}
\usepackage{color}

\usepackage{amsmath} 
\usepackage{multirow}
\usepackage{booktabs}
\usepackage{marvosym}

\begin{document}
\title{Structural Entities Extraction and Patient Indications Incorporation for Chest X-ray Report Generation}
\titlerunning{\textbf{S}tructural \textbf{E}ntities Extraction and Patient Indications \textbf{I}ncorporation (SEI)}

%
%

\author{Kang Liu\inst{1, 2, 3} \and
Zhuoqi Ma\inst{1, 2, 3, 4} \textsuperscript{(\Letter)} \and
Xiaolu Kang \inst{1,2,3} \and
Zhusi Zhong \inst{5} \and
Zhicheng Jiao\inst{4} \and
Grayson Baird \inst{4} \and
Harrison Bai \inst{6} \and
Qiguang Miao \inst{1,2,3}}

\authorrunning{K. Liu et al.}
%
\institute{School of Computer Science and Technology, Xidian University, Xi'an, China \and
Xi'an Key Laboratory of Big Data and Intelligent Vision, Xi'an, China \and
Key Laboratory of Collaborative Intelligence Systems, Ministry of Education, Xidian University, Xi'an, China \and 
Warren Alpert Medical School, Brown University, Providence, USA \and
School of Electronic Engineering, Xidian University, Xi'an, China \and
Department of Radiology and Radiological Sciences, Johns Hopkins University School of Medicine, Baltimore, USA\\
\email{zhuoqi\_ma@hotmail.com}}
\maketitle              
\begin{abstract}

The automated generation of imaging reports proves invaluable in alleviating the workload of radiologists. A clinically applicable reports generation algorithm should demonstrate its effectiveness in producing reports that accurately describe radiology findings and attend to patient-specific indications. In this paper, we introduce a novel method, \textbf{S}tructural \textbf{E}ntities extraction and patient indications \textbf{I}ncorporation (SEI) for chest X-ray report generation. Specifically, we employ a structural entities extraction (SEE) approach to eliminate presentation-style vocabulary in reports and improve the quality of factual entity sequences. This reduces the noise in the following cross-modal alignment module by aligning X-ray images with factual entity sequences in reports, thereby enhancing the precision of cross-modal alignment and further aiding the model in gradient-free retrieval of similar historical cases. Subsequently, we propose a cross-modal fusion network to integrate information from X-ray images, similar historical cases, and patient-specific indications. This process allows the text decoder to attend to discriminative features of X-ray images, assimilate historical diagnostic information from similar cases, and understand the examination intention of patients. This, in turn, assists in triggering the text decoder to produce high-quality reports. Experiments conducted on MIMIC-CXR validate the superiority of SEI over state-of-the-art approaches on both natural language generation and clinical efficacy metrics.

\keywords{Chest X-ray report generation  \and Structural entities extraction \and Patient-specific indications \and Cross-modal fusion \and Similar historical cases.}
\end{abstract}
\section{Introduction}
Radiology reports play a crucial role in delivering clear, accurate, and easily understandable medical information, thereby facilitating effective communication between doctors and patients. Nevertheless, this task is both highly specialized and time-consuming. Additionally, variations in proficiency, experience, and individual habits among radiologists would inevitably impact the quality and consistency of reports. Fortunately, the rapid evolution of artificial intelligence techniques, particularly deep learning \cite{devlin-etal-2019-bert,he-resnet,vaswani-transformer}, has significantly propelled the advancement of chest X-ray report generation (CXRG).

Currently, there is a considerable body of studies on medical report generation \cite{chen-etal-2020-generating,zhang-tandemnet,hou-miccai,wang2023metransformer,yulong-miccai}, contributing significantly to the improvement of clinical effectiveness and linguistic fluency in reports. However, two challenges persist in CXRG: 1) To truly meet clinical needs, report generation processes should incorporate patient-specific indications, such as previous treatment history or responses to specific diagnostic requirements, which cannot be derived exclusively from medical images. 2) Existing methods face challenges in effectively focusing on the cross-modal alignment between medical images and reports. This is attributed to the practice of assigning equal weights to presentation-style elements (e.g., sentence structure and grammar) and factual vocabulary (e.g., findings) in reports. Unfortunately, this limitation impacts their clinical efficacy. 

In response to the challenges above, previous studies have made specific attempts. Concerning challenge 1): \cite{access-indication} and \cite{tian-miccai-indication} leverage BiLSTM to encode the indication section with specific terms, facilitating the generation of purposeful reports. \cite{ml4h-indication-rg} adopts the LLaMA model \cite{llama} to generate reports based on indications and predicted positive conditions, falling short in fully exploiting valuable information within medical images. Furthermore, many existing works \cite{chen-etal-2021-cross-modal,kong-transq-miccai,yuan-multi-view-miccai} treat the CXRG task as an image-to-text generation problem, neglecting the effect of patient-specific indications on CXRG. Regarding challenge 2): Numerous studies directly utilize reports and medical images for cross-modal alignment at various granularities (e.g., instance-level \cite{yang-m2kt,zhang-kad}, sentence-level \cite{cheng-prior}, and token-level \cite{wang-mgca}). However, these methods treat presentation-style elements and factual vocabulary equally in reports, potentially affecting the quality of cross-modal alignment and, consequently, the clinical efficacy of the generated reports. In light of this limitation, building directly upon the outcomes of RadGraph \cite{jain-radgraph}, KAD \cite{zhang-kad} achieves instance-wise cross-modal alignment between images and the factual vocabulary in reports, while \cite{yan2023style} focuses on learning the mapping relationship between them using an encoder-decoder framework. Nevertheless, both methods overlook the noise and redundancy in the RadGraph outcomes.

In this paper, we introduce a novel method, \textbf{S}tructural \textbf{E}ntities extraction and patient indications \textbf{I}ncorporation (SEI), for chest X-ray report generation. SEI involves two stages: training the cross-modal alignment module and training the report generation module. In the first stage, SEI employs the structural entities extraction (SEE) approach to eliminate presentation-style vocabulary in reports and enhance the quality of factual entity sequences. Subsequently, a cross-modal alignment module is introduced between X-ray images and factual entity sequences in reports, ensuring that the extracted image features implicitly preserve semantic similarity with their corresponding reference reports. In the second stage, leveraging the pre-trained model from the first stage, we conduct a gradient-free retrieval of similar historical cases for each sample from the training set. Following this, the cross-modal fusion network is deployed to integrate these cases, patient-specific indications, and imaging information. This process enables the text decoder to assimilate historical diagnostic information from similar cases, understand the examination intention of patients (e.g., symptoms), and attend to discriminative features of X-ray images. Finally, this contributes to triggering the text decoder for the generation of high-quality reports. The effectiveness of our proposed method is successfully validated on MIMIC-CXR in both specific and general scenarios, outperforming multiple state-of-the-art methods. 

In summary, our key contributions are as follows: 1) We develop a structural entities extraction approach to extract factual entity sequences from reports. This step reduces the noise in the cross-modal alignment process, facilitating gradient-free retrieval of similar historical cases from the training set. 2) We introduce a cross-modal fusion network to integrate similar historical cases, patient-specific indications, and imaging information. This allows the text decoder to attend to discriminative features of X-ray images, assimilate historical empirical information from similar cases, and understand the examination intention of patients. 3) Experiments on MIMIC-CXR demonstrate that our SEI achieves state-of-the-art performance across almost all metrics. This highlights the capability of our model to generate reports with encouraging clinical efficacy and linguistic fluency. 

\section{Method}
\begin{figure}[!t]
    \centering
    \includegraphics[width=1\linewidth]{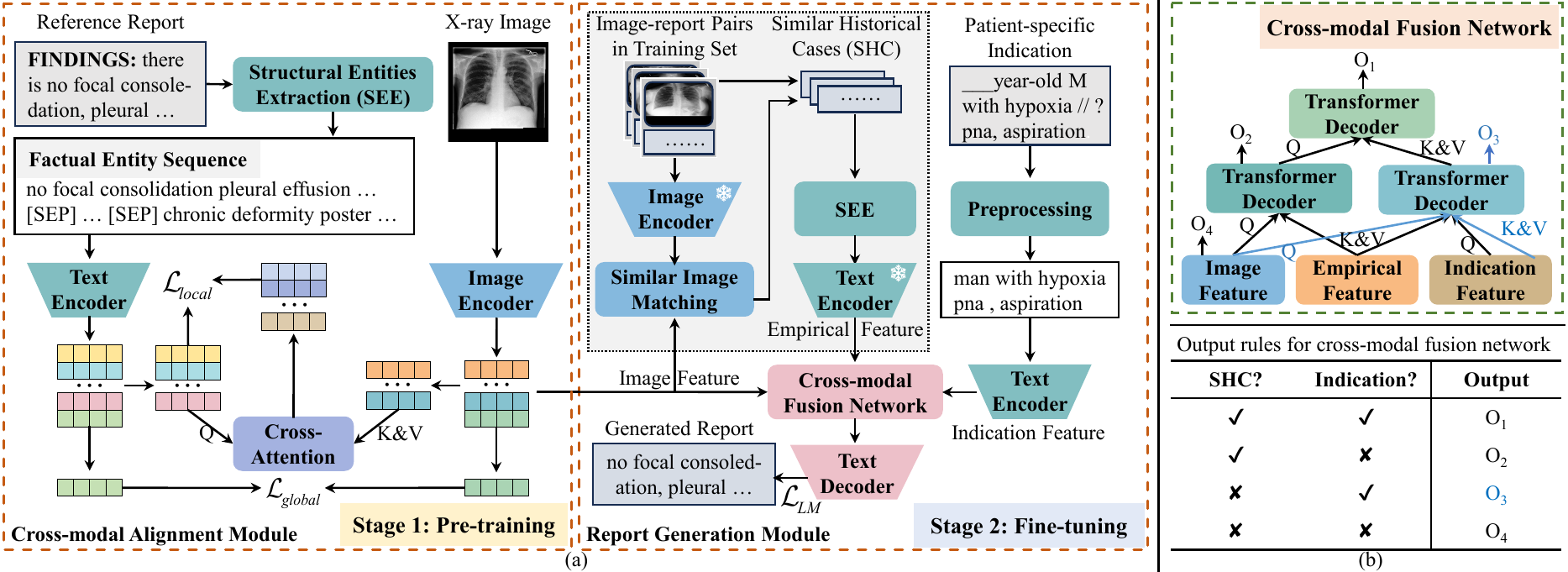}
    \caption{Illustration of our SEI and cross-modal fusion network. (a) Overview of SEI, featuring dual encoders for extracting uni-modal features and a text decoder for report generation using X-ray images, similar historical cases (SHC), and patient-specific indications. The training paradigm of SEI includes 1) pre-training via the cross-modal alignment module; 2) gradient-free retrieval of similar historical cases using the pre-trained model from step 1), shown in the light grey box; 3) fine-tuning using the report generation module.  (b) Details and output rules of the cross-modal fusion network.}
    \label{fig:1}
\end{figure}

Our objective is to train a model capable of generating a report for a given X-ray image, conditioned on similar historical cases and patient-specific indications. As shown in Fig.~\ref{fig:1}, the model comprises two stages: pre-training,  which involves a cross-modal alignment module enhanced by factual entity sequences, and fine-tuning, which is a report generation module based on both similar historical cases and patient-specific indications. 

\subsection{Cross-modal Alignment Module Enhanced by Factual Entity Sequences} \label{section: cross-modal alignment}
\textbf{Structural entities extraction approach for extracting factual entity sequences.} Motivated by \cite{yan2023style}, we devise the structural entities extraction (SEE) approach. More precisely, we first extract entities from reports using RadGraph \cite{jain-radgraph}. Afterward, we remove entities spanning two sentences, for example, “\textit{in place . Swan Ganz}”. For redundant entities at the same location, such as “\textit{1.9 \(\times \) 1.0 cm}” and “\textit{1.0 cm}”, we retain only the longest entity, specifically, the former. Following this, all remaining entities in the report are organized in their original order and divided into multiple factual entity subsequences, such as “\textit{AICD in place}”, using sentence periods. Notably, when a factual entity subsequence contains an “\textit{OBS-DA}” (or “\textit{OBS-U}”) entity \cite{jain-radgraph} in a factual entity subsequence, the keyword “\textit{no}” (or “\textit{maybe}”) is added at the beginning of the subsequence. Finally, we employ the [SEP] token to connect these subsequences, creating factual entity sequences. SEE eliminates the presentation-style elements in reports and enhances the quality of factual entity sequences, reducing the noise in the following cross-modal alignment module. 

\noindent \textbf{Cross-modal alignment between X-ray images and factual entity sequences.} To maintain consistency in representations of the same instance across different modalities, we adopt the PRIOR \cite{cheng-prior} method to define global image-to-report alignment loss \({\cal L}_{global}^{R \leftarrow I}\) and global report-to-image alignment loss \({\cal L}_{global}^{I \leftarrow R}\). Recognizing the importance of fine-grained features in medical report generation, we utilize the PRIOR \cite{cheng-prior} method to define local image-to-report alignment loss \({\cal L}_{local}\). Note that our approach differs from the PRIOR method in that we extract text features from factual entity sequences in reports rather than the original reports. This allows our model to focus on the cross-modal alignment between medical images and factual entity sequences in reports. To summarize, the training objective for the cross-modal alignment module is \({\cal L}_{global}^{R \leftarrow I}+{\cal L}_{global}^{I \leftarrow R}+{\cal L}_{local}\).

\subsection{Report Generation Module Based on Both Similar Historical Cases and Patient-specific Indications}
A doctor formulates a particular treatment based on both patient-specific indications (i.e., the examination intention of a patient) and previous patients with similar clinical findings or symptoms. Consequently, we introduce the report generation module based on both similar historical cases and patient-specific indications. In this Section, we will describe similar historical cases retrieval, cross-modal fusion network, and the report generation module.

\noindent \textbf{Similar historical cases retrieval.} Building on the pre-trained model from Section~\ref{section: cross-modal alignment}, we begin by extracting aligned image features. Given that these features implicitly preserve semantic similarity with their corresponding reference reports, we utilize a similar image matching approach (e.g., the dot product) to conduct gradient-free retrieval of similar historical cases for each sample from the training set. To enhance retrieval efficiency, we employ the Faiss tool \cite{johnson2019billion} to compute the similarity between image features.

\noindent \textbf{Cross-modal fusion network.} To integrate information from similar historical cases, patient-specific indications, and X-ray images, we propose the cross-modal fusion network. The network comprises three Transformer Decoder layers \cite{chen-ptunifier,chen2022align}, each featuring a self-attention sub-layer, cross-attention sub-layer, and feed-forward sub-layer. As shown in Fig.~\ref{fig:1}(b), the network automatically extracts information related to X-ray images and patient-specific indications from similar historical cases, respectively, enriching image and indication features. Subsequently, it further integrates these enriched features through a Transformer Decoder layer. Notably, even if some samples lack indications, the proposed cross-modal fusion network can fully utilize the available indication information through the output rules shown in Fig. \ref{fig:1}(b). These processes allow the text decoder to attend to discriminative features of X-ray images, assimilate historical diagnostic information from similar cases, and understand the examination intention of patients.

\noindent \textbf{Report generation module.} We initialize the image and text encoders with the pre-trained model discussed in Section~\ref{section: cross-modal alignment}. Afterward, we preprocess patient-specific indications. Specifically, we delete illegal characters (e.g., “/”, “\textit{\_}”, “\textit{@}”) and invalid words (e.g., “\textit{history:}”, “\textit{-year-old}”, “\textit{year old}”). When gender information is present in a patient-specific indication, we unify it as either “\textit{man}” or “\textit{woman}”. Finally, the report generation module is optimized by minimizing the negative log-likelihood \( {P}\left( {\left. {\tilde y_t^i} \right|{\boldsymbol{X}^i},c_K^i,I^i,\tilde y_{j,j<t}^i} \right)\): 
\begin{equation}
    {{{\cal L}}_{LM}} = - \frac{1}{B}\sum\limits_{i = 1}^B {\sum\limits_{t = 1}^M {\log {{P}\left( {\left. {\tilde y_t^i} \right|{\boldsymbol{X}^i},c_K^i,I^i,\tilde y_{j,j<t}^i} \right)}}} ,
\end{equation}
where \textit{B}, \textit{M}, \textbf{\textit{X}}, \(c_K\), \textit{I}, and \(\tilde y_{j,j<t}\) denote the batch size, the maximum length of tokens generated by the text decoder, image features extracted by the image encoder, the set with \textit{K} similar historical cases, the patient-specific indication, and the word sequence predicted by the text decoder for the first \(t-1\) time steps, respectively. 

\section{Experiments}
\subsection{Datasets, Evaluation Metrics, and Experimental Settings}
\textbf{Datasets.} We evaluate the effectiveness of our SEI using the MIMIC-CXR\footnote{https://physionet.org/content/mimic-cxr/2.0.0/} \cite{johnson-mimic-cxr-jpg} dataset, following the official partitioning settings.  Our approach aligns with prior studies \cite{chen-etal-2020-generating,li-dcl,tanida-rgrg,yang-gsket}, utilizing the findings section of raw radiology reports as reference reports. Additionally, we filter out samples with either empty or clinically meaningless report content, such as “\emph{Portable supine chest radiograph\_\_at 23:16 is subnitted.}”. Therefore, the training, validation, and test sets include 269,239 (150,957), 2,113 (1,182), and 3,852 (2,343) chest X-ray images (reports), respectively. All reproducibility methods utilize the same test set to ensure a fair and consistent comparison.

\begin{table}[!t]
\caption{Comparison of our SEI with  SOTA approaches on MIMIC-CXR. \(^{\dagger}\) means quoted results from the published literature, excluding RG and CX5, as these were not calculated in the literature. The remaining results are reproduced using the official code and checkpoints. The best values for each \({{M_{gt}}}\) are highlighted in \textbf{bold}, with the second-best values in \underline{underlined}. SEI-\textit{n} denotes our SEI incorporated with information from \textit{n} similar historical cases. Larger values for each metric indicate better performance.}
\label{table: main_results}
\centering
\begin{tabular}{c|c|cccc|ccc} 
\toprule
\multirow{2}{*}{\textbf{Method}}       & \multirow{2}{*}{\({\boldsymbol{M_{gt}}}\)} & \multicolumn{4}{c|}{\textbf{NLG Metrics}}                                                             & \multicolumn{3}{c}{\textbf{CE Metrics}}                   \\
                                       &                      & \textbf{BL-2}         & \textbf{BL-4}         & \textbf{MTR}          & \textbf{R\_L}         & \textbf{RG}    & \textbf{CX5}  & \textbf{CX14}  \\ 
\midrule
\multirow{2}{*}{R2Gen \cite{chen-etal-2020-generating} (EMNLP’20)}      & 100{\(^{\dagger}\)}                  & 0.218                 & 0.103                 & 0.137                 & 0.264                 & 0.207          & 0.340          & 0.340           \\
                                       & \textit{Cpl.}        & 0.209                 & 0.097                 & 0.135                 & 0.266                 & 0.211          & 0.339          & 0.338           \\ 
\hline
\multirow{2}{*}{R2GenCMN \cite{chen-etal-2021-cross-modal} (ACL’21)}     & 100{\(^{\dagger}\)}                  & 0.218                 & 0.106                 & 0.142                 & 0.278                 & 0.220          & 0.461          & 0.278           \\
                                       & \textit{Cpl.}        & 0.198                 & 0.090                 & 0.133                 & 0.268                 & 0.223          & 0.464          & 0.393           \\ 
\hline
GSKET \cite{yang-gsket} (MedIA’22)                       & 80{\(^{\dagger}\)}                   & 0.228                 & 0.115                 & -                     & 0.284                 & -              & -              & 0.371           \\ 
\hline
\multirow{2}{*}{CGPT2 \cite{nicolson-improving} (ARTMED’23)}     & 60{\(^{\dagger}\)}                   & \underline{0.248} & 0.127                 & 0.155                 & 0.286                 & 0.223          & 0.463          & 0.391           \\
                                       & \textit{Cpl.}        & 0.204                 & 0.102                 & 0.138                 & 0.277                 & 0.237          & 0.483          & 0.434           \\ 
\hline
\multirow{2}{*}{M2KT \cite{yang-m2kt} (MedIA’23)}       & 80{\(^{\dagger}\)}                   & 0.237                 & 0.111                 & 0.137                 & 0.274                 & 0.204          & 0.477          & 0.352           \\
                                       & \textit{Cpl.}        & 0.204                 & 0.085                 & 0.133                 & 0.244                 & 0.210          & 0.483          & 0.413           \\ 
\hline
DCL \cite{li-dcl} (CVPR’23)                        & 90{\(^{\dagger}\)}                   & -                     & 0.109                 & 0.150                 & 0.284                 & -              & -              & 0.373           \\ 
\hline
RGRG \cite{tanida-rgrg} (CVPR’23)                       & \textit{Cpl.}{\(^{\dagger}\)}        & \textbf{0.249}        & \underline{0.126}         & \textbf{0.168}        & 0.264                 & -              & \textbf{0.547} & 0.447           \\ 
\hline
\multirow{5}{*}{\textbf{SEI-0 (ours)}} & 60                   & \textbf{0.268}        & \underline{0.146} & \underline{0.164} & \underline{0.300} & \textbf{0.239} & \underline{0.505}  & \underline{0.437}   \\
                                       & 80                   & \underline{0.250} & \underline{0.135} & \underline{0.158} & \textbf{0.300}        & \textbf{0.250} & \underline{0.531}  & \underline{0.452}   \\
                                       & 90                   & \underline{0.244}         & \underline{0.131}         & \underline{0.156}         & \underline{0.299}         & \textbf{0.252} & \underline{0.536}  & \underline{0.455}   \\
                                       & 100                  & \underline{0.240} & \underline{0.129} & \underline{0.154} & \underline{0.298} & \textbf{0.252} & \underline{0.539}  & \underline{0.457}   \\
                                       & \textit{Cpl.}        & 0.231                 & 0.123                 & 0.150                 & \textbf{0.297}        & \textbf{0.252} & 0.541          & \underline{0.457}   \\ 
\hline
\multirow{5}{*}{\textbf{SEI-1 (ours)}} & 60                   & \textbf{0.268}        & \textbf{0.148}        & \textbf{0.167}        & \textbf{0.301}        & \underline{0.236}  & \textbf{0.509} & \textbf{0.445}  \\
                                       & 80                   & \textbf{0.257}        & \textbf{0.140}        & \textbf{0.162}        & \textbf{0.300}        & \underline{0.247}  & \textbf{0.535} & \textbf{0.457}  \\
                                       & 90                   & \textbf{0.251}        & \textbf{0.137}        & \textbf{0.160}        & \textbf{0.300}        & \underline{0.248}  & \textbf{0.539} & \textbf{0.459}  \\
                                       & 100                  & \textbf{0.247}        & \textbf{0.135}        & \textbf{0.158}        & \textbf{0.299}        & \underline{0.249}  & \textbf{0.542} & \textbf{0.460}  \\
                                       & \textit{Cpl.}        & \underline{0.238}         & \textbf{0.128}        & \underline{0.154}         & \underline{0.296}         & \underline{0.249}  & \underline{0.545}  & \textbf{0.460}  \\
\bottomrule
\end{tabular}
\end{table}

\noindent \textbf{Evaluation Metrics.} We utilize metrics for both conventional natural language generation (NLG) and clinical efficacy (CE) to estimate lexical similarity and clinical effectiveness between generated and reference reports. Specifically, NLG metrics include BLEU-2 (BL-2), BLEU-4 (BL-4), METEOR (MTR), and ROUGE\_L (R\_L), calculated using pycocoevalcap\footnote{https://github.com/tylin/coco-caption}. CE metrics comprise F\textsubscript{1,mic-14} CheXbert (CX14), F\textsubscript{1,mic-5} CheXbert (CX5) \cite{smit-etal-2020-chexbert}, and F\textsubscript{1} RadGraph (RG) \cite{delbrouck-etal-2022-improving,jain-radgraph}, calculated by  f1chexbert\footnote{https://pypi.org/project/f1chexbert/} and radgraph\footnote{https://pypi.org/project/radgraph/}, respectively. 

\noindent \textbf{Experimental Settings.} We regard ResNet101 \cite{chen-etal-2020-generating,he-resnet} pre-trained on ImageNet as the image encoder, and a six-layer pre-trained SciBERT \cite{Beltagy2019} model as the text encoder. In addition, we adopt the memory-driven Transformer, designed by R2Gen \cite{chen-etal-2020-generating}, as the text decoder and train it from scratch. In the first stage (i.e., training the cross-modal alignment module), we employ the AdamW optimizer with an initial learning rate of 5e-5, conducting training for 100 epochs. In the second stage (i.e., training the report generation module), we utilize the RAdam optimizer with a learning rate of 5e-5 for 30 epochs. The optimal model is selected based on cumulative scores, considering RG, CX14, and BL-4 metrics on the validation set. Subsequently, we present the results on the test set accordingly. 

\begin{table}[!t]
\centering
\caption{Ablation studies on MIMIC-CXR in the general scenario. SEI-\textit{n} represents our SEI incorporated with information from \textit{n} similar historical cases. The best result is indicated in \textbf{bold}. SHC denotes similar historical cases, and w/o is without.}
\label{table: ablation_study}
\begin{tabular}{c|l|cccc|ccc} 
\toprule
\multirow{2}{*}{\textbf{Settings}} & \multirow{2}{*}{\textbf{Model}}  & \multicolumn{4}{c|}{\textbf{NLG Metrics}}                                 & \multicolumn{3}{c}{\textbf{CE Metrics}}                   \\ 
                                   &                                  & \textbf{BL-2}   & \textbf{BL-4}   & \textbf{MTR}   & \textbf{R\_L}   & \textbf{RG}    & \textbf{CX5}   & \textbf{CX14}   \\ 
\midrule
(a)                                & Base (R2Gen \cite{chen-etal-2020-generating})& 0.209          & 0.097          & 0.135          & 0.266          & 0.211          & 0.339          & 0.338           \\ 
\hline
(b)                                & (a)\(+\)cross-modal
  module& 0.206          & 0.098          & 0.138          & 0.277          & 0.234          & 0.513          & 0.431           \\ 
\hline
(c)                                & SEI-1 w/o indications& 0.228          & 0.109          & 0.148          & 0.279          & 0.241          & 0.542          & \textbf{0.474}  \\ 
\hline
(d)                                & SEI-1 w/o SHC (\textbf{SEI-0})& 0.231          & 0.123          & 0.150          & \textbf{0.297} & \textbf{0.252} & 0.541          & 0.457           \\ 
\hline
(e)                                & \textbf{SEI-1}& \textbf{0.238} & \textbf{0.128} & \textbf{0.154} & 0.296          & 0.249          & \textbf{0.545} & 0.460           \\
\bottomrule
\end{tabular}
\end{table}

\subsection{Main Results}
To comprehensively assess coherence and integrality between generated and reference reports, we preserve the generated reports unaltered and truncate reference reports to a specific length, denoted as \(M_{gt}\), to establish ground truth.  This allows us to evaluate the model performance in different scenarios. In specific scenarios (i.e., \({M_{gt}} \in \left\{ {60,80,90,100} \right\}\)), such as emergency diagnoses, concise medical reports prove more effective.  In contrast, comprehensive and detailed reports are essential in a general scenario (i.e., \(M_{gt}=Cpl.\), where \textit{Cpl.} represents the length of complete reference reports). We compare our SEI with seven state-of-the-art (SOTA) approaches: R2Gen \cite{chen-etal-2020-generating}, R2GenCMN \cite{chen-etal-2021-cross-modal}, GSKET \cite{yang-gsket}, CvT2DistillGPT2 (CGPT2) \cite{nicolson-improving}, M2KT \cite{yang-m2kt}, DCL \cite{li-dcl}, and RGRG \cite{tanida-rgrg}. 

Results on MIMIC-CXR are presented in Table~\ref{table: main_results}. Upon observation, our approach consistently outperforms previous state-of-the-art methods across diverse scenarios, achieving significant improvements on almost all metrics, notably with a notable RG score of 0.252. These experimental findings highlight the capability of our SEI to generate reports with impressive clinical efficacy and linguistic fluency in various scenarios.

\subsection{Ablation Study}
Table~\ref{table: ablation_study} illustrates the positive effects of each component on model performance, particularly the similar historical cases and patient-specific indications. We observe from (c) and (d) that integrating them individually into the model leads to significant improvements in both NLG and CE metrics. Although (e), which integrates two components simultaneously, shows performance degradation compared to (c) and (d) on certain metrics, it enhances overall performance by 4.9\% and 1.9\% across all metrics, respectively. This may be attributed to potential interference between the components, hindering the full exploitation of their respective strengths. In addition, the absence of indications in some samples has resulted in an unstable fusion feature space. Addressing this issue remains a topic for future research.

\subsection{Qualitative Analysis}
\begin{figure}[!t]
    \centering
    \includegraphics[width=1\linewidth]{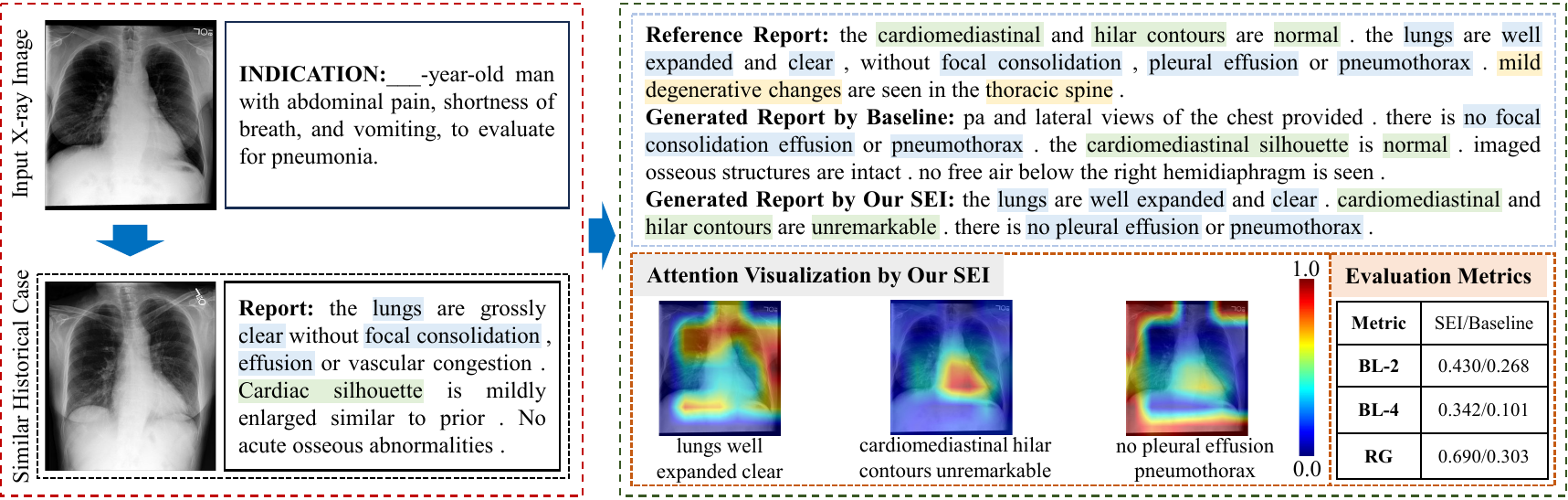}
    \caption{An example of generated reports and attention visualization on MIMIC-CXR test set. Distinct colors in the reference report indicate the factual entity subsequence within different sentences. Generated reports and similar historical cases are highlighted in matching colors. “\textit{Baseline}” represents the CGPT2 \cite{nicolson-improving} method.}
    \label{fig:2}
\end{figure}

The left side of Fig.~\ref{fig:2} presents the patient-specific indication and historical similar cases of the example, while the right side illustrates the generated reports and attention visualization. We observed that our generated report aligns with the phrase “\textit{to evaluate for pneumonia}” in the indication section and demonstrates a high level of consistency with the reference report in terms of clinical efficacy.

\section{Conclusion}
This paper introduced a novel method, \textbf{S}tructural \textbf{E}ntities extraction and patient indications \textbf{I}ncorporation (SEI), for chest X-ray report generation. SEI first developed the structural entities extraction approach to extract factual entity sequences from medical reports. This reduces the noise in the following cross-modal alignment module, thereby further aiding the model in gradient-free retrieval of similar historical cases from the training set. Subsequently, we proposed a cross-modal fusion network to integrate the information from X-ray images, similar historical cases, and patient-specific indications, ensuring the text decoder attended to discriminative features of X-ray images, assimilated historical empirical information from similar cases, and understood the examination intention of patients. Experiments on MIMIC-CXR in various scenarios proved that our SEI outperformed previous state-of-the-art methods. The ablation study demonstrated the significance of the structural entity extraction approach for the cross-modal alignment module, along with similar historical cases and patient-specific indications for the report generation module. While SEI exhibited impressive performance, it did not incorporate patient-specific indications during the cross-modal alignment phase. This aspect will be explored in future work.

\section*{Acknowledgments}
The work was jointly supported by the National Science and Technology Major Project under grant No. 2022ZD0117103, the National Natural Science Foundations of China under grant No. 62272364, the provincial Key Research and Development Program of Shaanxi under grant No. 2024GH-ZDXM-47, the Research Project on Higher Education Teaching Reform of Shaanxi Province under grant No. 23JG003, and the High-Performance Computing Platform of Xidian University.

%
%
%
\bibliographystyle{splncs04}
\bibliography{refs}

\begin{thebibliography}{10}
\providecommand{\url}[1]{\texttt{#1}}
\providecommand{\urlprefix}{URL }
\providecommand{\doi}[1]{https://doi.org/#1}

\bibitem{Beltagy2019}
Beltagy, I., Lo, K., Cohan, A.: Scibert: A pretrained language model for scientific text. In: EMNLP. pp. 3615--3620 (2019). \doi{10.18653/v1/D19-1371}

\bibitem{chen-ptunifier}
Chen, Z., Diao, S., Wang, B., Li, G., Wan, X.: Towards unifying medical vision-and-language pre-training via soft prompts. In: ICCV. pp. 23346--23356 (2023). \doi{10.1109/ICCV51070.2023.02139}

\bibitem{chen2022align}
Chen, Z., Li, G., Wan, X.: Align, reason and learn: Enhancing medical vision-and-language pre-training with knowledge. In: ACMMM. pp. 5152--5161. Association for Computing Machinery (2022). \doi{10.1145/3503161.3547948}

\bibitem{chen-etal-2021-cross-modal}
Chen, Z., Shen, Y., Song, Y., Wan, X.: Cross-modal memory networks for radiology report generation. In: ACL. vol.~1, pp. 5904--5914 (2021). \doi{10.18653/v1/2021.acl-long.459}

\bibitem{chen-etal-2020-generating}
Chen, Z., Song, Y., Chang, T.H., Wan, X.: Generating radiology reports via memory-driven transformer. In: EMNLP. pp. 1439--1449 (2020). \doi{10.18653/v1/2020.emnlp-main.112}

\bibitem{cheng-prior}
Cheng, P., Lin, L., Lyu, J., Huang, Y., Luo, W., Tang, X.: Prior: Prototype representation joint learning from medical images and reports. In: ICCV. pp. 21361--21371 (2023). \doi{10.1109/ICCV51070.2023.01953}

\bibitem{delbrouck-etal-2022-improving}
Delbrouck, J.B., Chambon, P., Bluethgen, C., Tsai, E., Almusa, O., Langlotz, C.: Improving the factual correctness of radiology report generation with semantic rewards. In: EMNLP. pp. 4348--4360 (2022). \doi{10.18653/v1/2022.findings-emnlp.319}

\bibitem{devlin-etal-2019-bert}
Devlin, J., Chang, M.W., Lee, K., Toutanova, K.: Bert: Pre-training of deep bidirectional transformers for language understanding. In: NAACL. vol.~1, pp. 4171--4186 (2019). \doi{10.18653/v1/N19-1423}

\bibitem{he-resnet}
He, K., Zhang, X., Ren, S., Sun, J.: Deep residual learning for image recognition. In: CVPR. pp. 770--778 (2016). \doi{10.1109/CVPR.2016.90}

\bibitem{hou-miccai}
Hou, Z., Yan, R., Wang, Q., Lang, N., Zhou, X.: Diversity-preserving chest radiographs generation from reports in one stage. In: MICCAI. vol. 14224, pp. 482--492 (2023). \doi{10.1007/978-3-031-43904-9\_47}

\bibitem{access-indication}
Huang, X., Yan, F., Xu, W., Li, M.: Multi-attention and incorporating background information model for chest x-ray image report generation. IEEE Access  \textbf{7},  154808--154817 (2019). \doi{10.1109/ACCESS.2019.2947134}

\bibitem{jain-radgraph}
Jain, S., Agrawal, A., Saporta, A., Truong, S., Duong, D.N.D.N., Bui, T., Chambon, P., Zhang, Y., Lungren, M., Ng, A., Langlotz, C., Rajpurkar, P., Rajpurkar, P.: Radgraph: Extracting clinical entities and relations from radiology reports. In: NeurIPS. vol.~1 (2021)

\bibitem{johnson-mimic-cxr-jpg}
Johnson, A.E., Pollard, T.J., Greenbaum, N.R., Lungren, M.P., Deng, C.y., Peng, Y., Lu, Z., Mark, R.G., Berkowitz, S.J., Horng, S.: Mimic-cxr-jpg, a large publicly available database of labeled chest radiographs. arXiv preprint arXiv:1901.07042  (2019)

\bibitem{johnson2019billion}
Johnson, J., Douze, M., Jégou, H.: Billion-scale similarity search with gpus. IEEE Transactions on Big Data  \textbf{7}(3),  535--547 (2019). \doi{10.1109/TBDATA.2019.2921572}

\bibitem{kong-transq-miccai}
Kong, M., Huang, Z., Kuang, K., Zhu, Q., Wu, F.: Transq: Transformer-based semantic query for medical report generation. In: MICCAI. vol. 13438, pp. 610--620 (2022). \doi{10.1007/978-3-031-16452-1\_58}

\bibitem{li-dcl}
Li, M., Lin, B., Chen, Z., Lin, H., Liang, X., Chang, X.: Dynamic graph enhanced contrastive learning for chest x-ray report generation. In: CVPR. pp. 3334--3343 (2023). \doi{10.1109/CVPR52729.2023.00325}

\bibitem{ml4h-indication-rg}
Nguyen, D., Chen, C., He, H., Tan, C.: Pragmatic radiology report generation. In: ML4H. vol.~225, pp. 385--402. PMLR (2023)

\bibitem{nicolson-improving}
Nicolson, A., Dowling, J., Koopman, B.: Improving chest x-ray report generation by leveraging warm starting. Artificial Intelligence in Medicine  \textbf{144},  102633 (2023). \doi{10.1016/j.artmed.2023.102633}

\bibitem{smit-etal-2020-chexbert}
Smit, A., Jain, S., Rajpurkar, P., Pareek, A., Ng, A., Lungren, M.: Combining automatic labelers and expert annotations for accurate radiology report labeling using {BERT}. In: EMNLP. pp. 1500--1519 (2020). \doi{10.18653/v1/2020.emnlp-main.117}

\bibitem{tanida-rgrg}
Tanida, T., Müller, P., Kaissis, G., Rueckert, D.: Interactive and explainable region-guided radiology report generation. In: CVPR. pp. 7433--7442 (2023). \doi{10.1109/CVPR52729.2023.00718}

\bibitem{tian-miccai-indication}
Tian, J., Zhong, C., Shi, Z., Xu, F.: Towards automatic diagnosis from multi-modal medical data. In: MICCAI. vol. 11797, pp. 67--74 (2019). \doi{10.1007/978-3-030-33850-3\_8}

\bibitem{llama}
Touvron, H., Lavril, T., Izacard, G., Martinet, X., Lachaux, M., Lacroix, T., Rozi{\`{e}}re, B., Goyal, N., Hambro, E., Azhar, F., Rodriguez, A., Joulin, A., Grave, E., Lample, G.: Llama: Open and efficient foundation language models. CoRR  \textbf{abs/2302.13971} (2023). \doi{10.48550/ARXIV.2302.13971}

\bibitem{vaswani-transformer}
Vaswani, A., Shazeer, N., Parmar, N., Uszkoreit, J., Jones, L., Gomez, A.N., Kaiser, L.u., Polosukhin, I.: Attention is all you need. In: NeurIPS. vol.~30 (2017)

\bibitem{wang-mgca}
Wang, F., Zhou, Y., Wang, S., Vardhanabhuti, V., Yu, L.: Multi-granularity cross-modal alignment for generalized medical visual representation learning. In: NeurIPS. vol.~35, pp. 33536--33549 (2022)

\bibitem{wang2023metransformer}
Wang, Z., Liu, L., Wang, L., Zhou, L.: Metransformer: Radiology report generation by transformer with multiple learnable expert tokens. In: CVPR. pp. 11558--11567 (2023). \doi{10.1109/CVPR52729.2023.01112}

\bibitem{yulong-miccai}
Xie, Y., Gu, L., Harada, T., Zhang, J., Xia, Y., Wu, Q.: Medim: Boost medical image representation via radiology report-guided masking. In: MICCAI. vol. 14220, pp. 13--23 (2023). \doi{10.1007/978-3-031-43907-0\_2}

\bibitem{yan2023style}
Yan, B., Liu, R., Kuo, D.E., Adithan, S., Reis, E.P., Kwak, S., Venugopal, V.K., O'Connell, C., Saenz, A., Rajpurkar, P., Moor, M.: Style-aware radiology report generation with radgraph and few-shot prompting. In: EMNLP. pp. 14676--14688 (2023). \doi{10.18653/v1/2023.findings-emnlp.977}

\bibitem{yang-m2kt}
Yang, S., Wu, X., Ge, S., Zheng, Z., Zhou, S.K., Xiao, L.: Radiology report generation with a learned knowledge base and multi-modal alignment. Medical Image Analysis  \textbf{86},  102798 (2023). \doi{10.1016/j.media.2023.102798}

\bibitem{yang-gsket}
Yang, S., Wu, X., Ge, S., Zhou, S.K., Xiao, L.: Knowledge matters: Chest radiology report generation with general and specific knowledge. Medical Image Analysis  \textbf{80},  102510 (2022). \doi{10.1016/j.media.2022.102510}

\bibitem{yuan-multi-view-miccai}
Yuan, J., Liao, H., Luo, R., Luo, J.: Automatic radiology report generation based on multi-view image fusion and medical concept enrichment. In: MICCAI. vol. 11769, pp. 721--729 (2019). \doi{10.1007/978-3-030-32226-7\_80}

\bibitem{zhang-kad}
Zhang, X., Wu, C., Zhang, Y., Xie, W., Wang, Y.: Knowledge-enhanced visual-language pre-training on chest radiology images. Nature Communications  \textbf{14}(1), ~4542 (2023). \doi{10.1038/s41467-023-40260-7}

\bibitem{zhang-tandemnet}
Zhang, Z., Chen, P., Sapkota, M., Yang, L.: Tandemnet: Distilling knowledge from medical images using diagnostic reports as optional semantic references. In: MICCAI. p. 320–328 (2017). \doi{10.1007/978-3-319-66179-7\_37}

\end{thebibliography}
%




\end{document}